\documentclass[aps,prb,twocolumn]{revtex4-1}
\usepackage{graphicx}
\usepackage{bm}

\begin{document}

\title{Comment on ``Transition from Bose glass to a condensate of
triplons in Tl$_{1-x}$K$_x$CuCl$_3$}
\author{A.~Zheludev, D.~H\"uvonen}
\affiliation{ Laboratory for Solid State Physics, ETH Z\"urich, Switzerland}

\begin{abstract}
We argue that the interpretation of the calorimetric data for
disordered quantum antiferromagnets Tl$_{1-x}$K$_x$CuCl$_3$ in terms
of Bose Glass physics by  F. Yamada {\it et al.} in [Phys. Rev. B
{\bf 83}, 020409(R) (2011)] is not unambiguous. A consistent
analysis shows no difference in the crossover critical index for the
disorder-free TlCuCl$_3$ and its disordered derivatives.
Furthermore, we question the very existence of a proper
field-induced thermodynamic phase transition in
Tl$_{1-x}$K$_x$CuCl$_3$.
\end{abstract}


\maketitle

The main claim of the paper by F. Yamada {\it et al.} is that K-substitution in
Tl$_{1-x}$K$_x$CuCl$_3$ changes the field-induced Mott Insulator
(MI) to BEC (Bose-Einstein Condensate) transition in the pure system
to a Bose Glass (BG) to BEC transition in the chemically disordered
compound. This conclusion is primarily drawn from a comparison of
the crossover exponents $\phi$. We question the validity of such an
analysis.

First, the data range is too small: at best it covers half a decade
in temperature, and as little as a quarter of a decade for some
cases. Second, the error of determining the peak position in field
scans is actually quite significant. The scattering of the raw data
that appears in the top three curves on Fig.~1b gives an estimate of
the noise in specific heat measurements: about $10^{-3}$~J/mol~K.
The method used for locating the maximum from noisy data is not
described in the paper. However, given the widths of the peaks, the
noise must translate into an error of determining the maximum
position of {\it at least} $\pm 0.1$~T at high temperature (as
admitted in the paper) and as much as $\pm 0.2$~T at low
temperatures. To illustrate the impact of these factors we have
re-analyzed the data digitized from Fig.~1c assuming an arbitrarily
fixed exponent $\phi=1.0$, and a very small measurement error of
0.05~T in the entire range. The fits yield almost perfect agreement
for {\it all three compositions}: $\chi^2=0.25$, $\chi^2=0.38$ and
$\chi^2=0.22$ for $x=0$, $x=0.22$ and $x=0.36$, respectively, and a
confidence of determination $R^2>0.985$ in all cases.

The rather small least-squares errors quoted for $\phi$ are actually
a poor measure of the confidence interval. In fact, $\phi$ has very
high covariance with another fit parameter, namely the critical
field. The latter is not measured independently. We found that the
covariance becomes particularly large when the fitting range is
restricted to low temperatures.

A meaningful comparison between the three compositions requires that
the same data range or the same decreasing fitting window procedure
is used in each case in a {\it consistent} analysis. The authors
obtain $\phi=1.53$ in the pure compound by including all data up to
2~K. Yet, the BG-to-BEC value $\phi=0.58$ in the $x=0.36$ material
is only found when the fitting range is restricted to below 1~K.
That such a free choice of fitting window can be misleading, is
illustrated by the following. Re-fitting digitized author's data for
$T>0.6$~K (with only four low-temperature points excluded) yields
MI-to-BEC-like exponents for {\it both} $x=0$ and $x=0.36$:
$\phi=1.74(0.2)$ and $\phi=1.39(17)$, respectively.  Thus, all
conclusions of the paper fully rely on {\it just these 4 data
points}.  The latter to correspond to the poorest-defined maxima.

A related issue concerns the $x=0.22$ data for which the authors
altogether avoid discussing the decreasing fitting window analysis.
Such an analysis of digitized data shows that $\phi$ remains close
to unity and exhibits no systematic decrease with decreasing
$T_{max}$. For $T_{max}=1$~K we get $\phi=1.02(0.3)$.

 The final
issue concerns the broad specific heat peaks in the raw $C(H)$ data.
A sharp lambda-anomaly that is a hallmark of a continuous phase
transition is entirely absent. It is conceivable that such peak
broadening is instrumental. Since the instrumental function is not
known, the transition point can not be unambiguously associated with
the measured maximum. The actual relation will depend on the
temperature dependence of the underlying specific heat divergence,
the measurement method and other factors. The peaks being broad as
they are, an 0.1~T mismatch at low temperatures can not be excluded.
The result would be a temperature-dependent systematic error that
directly affects $\phi$.

Alternatively, the specific heat of the samples actually {\it does
not diverge}. This implies that there is no continuous thermodynamic
phase transition, and that the correlation length remains finite.
Under these circumstances, any discussion of ``criticality'' or
critical indexes loses its significance. The absence of a true phase
transition may itself be a result of chemical disorder or due to a
staggered g-tensor and Dzyaloshinskii-Moriya interactions. \cite{DM}

Considering what is said above, it is not surprising that ``the BEC
phase is not reached even at the lowest temperature'' in ESR
experiments. What is surprising, is that it is at these very
temperatures (0.55~K) the authors claim to see evidence of the BEC
phase is bulk measurements.

This work is partially supported by the Swiss National Science
Foundation and by MANEP under Project 6.



\end{document}